\documentclass[a4paper,12pt]{article}
\pdfoutput=1
\usepackage{graphicx,subfigure,amsmath,amssymb,multirow}
\usepackage{cite}

\newlength{\dinwidth}
\newlength{\dinmargin}
\usepackage{color}

\begin{document}
  \title{Study of $B_{c}$$\to$$ J/\psi V $ and $B_{c}^{*}$$\to$$\eta_{c}V$ decays within the QCD factorization}
   \author{Qin Chang$^{a,b}$, Li-Li Chen$^{a}$ and Shuai Xu$^{a}$\\
{ $^a$\small Institute of Particle and Nuclear Physics, Henan Normal University, Henan 453007, China}\\
{ $^b$\small Institute of Particle Physics and Key Laboratory of Quark and Lepton Physics (MOE),}\\[-0.2cm]
{ \small Central China Normal University, Wuhan 430079, China}}
\date{}
 \maketitle

 \begin{abstract}
 In this paper, we study the non-leptonic $B_{c}\to J/\psi V$ and $B_{c}^{*}\to\eta_{c}V$~$(V=\rho\,,K^{*})$ weak decays in the framework of QCD factorization.
 In the evaluation, the form factors are calculated by using the Bauer-Stech-Wirbel model and the light-front  quark model, respectively.
 Besides the longitudinal amplitude, the power-suppressed transverse contributions are also evaluated at next-to-leading order. The predictions for the observables of $B_{c}\to J/\psi V$ and $B_{c}^{*}\to\eta_{c}V$ decays are presented.
 We find that the NLO QCD  contribution presents about $ 8\%$ correction to the branching ratios, and
the longitudinal polarization fractions of these decays are at the level of $(80\sim90)\%$.
In addition, we suggest direct measurements on  some useful ratios, $R_{K^{*}/\rho}^{(\lambda=0)}$ and $\widetilde{R}^{(\lambda=0)}_{K^{*}/\rho}$, which are very suitable for test the consistence between theoretical prediction and data because their theoretical uncertainties can be well-controlled.
    \end{abstract}
  \maketitle

  \newpage
  \section{Introduction}
  \label{sec01}
  The $B_{c}$ and $B_{c}^{*}$ mesons are the ground states of the $(b\bar c)$ system having the total angular moment and parity quantum number, $J^{P}=0^{-}$  and $1^{-}$, respectively~\cite{Patrignani:2016xqp}.
   The heavy $b\bar c$ state is first discovered by CDF collaboration via the semileptonic weak decay $B_c\to J/\psi \ell \bar{\nu}$ with $J/\psi$ decaying into muon pairs~\cite{Abe:1998wi}, and has soon attracted much attention due to its interesting properties.
   The $B_{c}$ and $B_{c}^{*}$ mesons are the ``double heavy-flavored'' binding systems, and share many features with the heavy quarkonia. In particular, their  weak decays provide windows for testing the predictions of the standard model~(SM) and can shed light on new physics scenarios.

    Because the mass of $B_{c}$ meson lies below the $BD$ threshold, $B_{c}$ meson cannot annihilate into gluons via strong interaction, and its decay is dominated by the weak interaction. As to the $B_{c}^{*}$ meson, Ref. \cite{Dowdall:2012ab} has made predictions for its mass that 
    $m_{B_{c}^{*}}-m_{B_{c}}\simeq 54\pm 3$ MeV, which is less than ${m_{\pi}}$.
  Moreover, the mass of $B_{c}^{*}$ lies below the $BD$ mesons threshold too.
    As a consequence, the $B_{c}^{*}$ meson also can not
   decay through the strong interaction but through the flavor-conserving electromagnetic transition
   and  flavor-changing weak transition.
 The radiative  decay mode $B_{c}^{*}\to B_{c}\gamma$ is the dominant progress,
  but is strongly suppressed by the compact phase space, which results in a very short lifetime,
   $\tau_{B_{c}^{*}}\sim {\cal O}(10^{-18}s)$~\cite{Simonis:2016pnh}, relative to the $B_{c}$ meson.

Experimentally, the production of $B_{c}$ and $B_{c}^*$  mesons in hadron collisions
is relatively rarer than the other $b$ mesons~\cite{Brambilla:2010cs}.
However, thanks to the rapid development of the experimental technology, their weak decays are still hopeful to be observed soon.
At the Large Hadron Collider (LHC) with a luminosity of about ${\cal L}=10^{34}{\rm cm}^{-2}{\rm s}^{-1}$,  around $5\times 10^{10}$ $B_c$ events can be produced  per year~\cite{Gouz:2002kk}, and the measurements of the mass and lifetime of  $B_{c}$ meson have reached a very precise degree, for instance, $m_{B_{c}}=6276.28\pm1.44\pm0.36$ MeV~\cite{Aaij:2013gia} and $\tau_{B_{c}}=513.4\pm11.0\pm5.7\,{\rm fs}$~\cite{Aaij:2014gka} reported by the LHCb collaboration. Benefiting from the large production rate at LHC, some $B_{c}$ decay channels have been observed by LHCb collaboration, for instance: the $B_{c}^{+}\to J/\Psi \pi^{+}\pi^{-}\pi^{+}$~\cite{LHCb:2012ag}, $\Psi(2S)\pi^{+}$~\cite{Aaij:2013oya}, $ J/\Psi D_{s}^{(*)}$~\cite{Aaij:2013gia}, $J/\Psi K^{+}$~\cite{Aaij:2013vcx}, $ J/\Psi K^{+} K^{-} \pi^{+}$~\cite{Aaij:2013gxa} and $D^{0} K^{+}$~\cite{Aaij:2017kea} decay modes and the ratio $\mathcal{B}(B_c^+\,\to\,J/\psi\tau^+\nu_\tau)$/$\mathcal{B}(B_c^+\,\to\,J/\psi\mu^+\nu_\mu)$~\cite{Aaij:2017tyk} induced by the $b$ quark decay, the first $c$ quark decay mode $B_{c}^{+} \to B_{s}^{0}\pi^{+}$~\cite{Aaij:2013cda} and the baryonic decay mode $B_{c}\to J/\Psi p\bar{p}\pi^{+}$~\cite{Aaij:2016xxs} {\it etc.}. In the near future, more $B_{c}$ weak decays are expected to be measured at LHC with its high collision energy, high luminosity and the  large production cross section~\cite{Chang:1994aw,Chang:1996jt,Chang:2003cr,Chang:2005wd}. In addition, some $B_{c}^{*}$ weak decays  are also possible to be observed in the future even though there is  no experimental information for now.

 Different from the other heavy mesons, since both of the constituents
 $(\bar b, c)$ in $B_c$  and $B_c^*$ mesons are heavy,  they can decay individually.
Generally, the decay modes can be divided into three types~\cite{Bigi:1995fs,Beneke:1996xe,9912424,Chang:2001jn,Chang:2000ac}:
  (1) the $b \to (c,u)W^{-}$ process with $\bar{c}$-quark as a spectator;
  (2) the $\bar{c} \to (\bar{s},\bar{d})W^{-}$ process with $b$-quark as a spectator;
  (3) the pure weak annihilation $b\bar{c} \to W^{-}$ transition.
    Therefore, $B_{c}$ and  $B_{c}^{*}$  mesons have abundant weak decay channels, and thus provide a fertile ground for testing the SM and studying the relevant dynamical mechanism, for instance, the perturbative and nonperturbative QCD, final state interactions and heavy quarkonium properties, {\it etc.}.  In the past years, some theoretical investigations have been carried out on the properties of $B_c$ meson decays based on the QCD-inspired approaches, for instance, the operator product expansion~\cite{Bigi:1995fs,Beneke:1996xe}, the QCD sum rule~\cite{Kiselev:1999sc,Kiselev:2001ej,Kiselev:2003ds,Kiselev:2000pp}, the nonrelativistic QCD~\cite{Bodwin:1994jh}, the naive factorization~(NF)~\cite{Kar:2013fna}, the pQCD factorization~\cite{Du:1991np,Yang:2010ba,Sun:2014ika,Sun:2014jka,Liu:2009qa,Rui:2015iia,Rui:2016opu}, the QCD factorization~(QCDF)~\cite{Sun:2015exa,DescotesGenon:2009ja,Wang:2016qli,Chang:2017ivy}, the QCD relativistic potential models~\cite{Colangelo:1999zn,Kiselev:2000jc} and the Bethe-Salpeter method~\cite{Ju:2014oha,Chang:2014jca}.

 In this paper, we will concentrate on the bottom changing weak decays,  $B_{c}$$\to$$ J/\psi V $ and $B_{c}^{*}$$\to$$\eta_{c}V$ $(V=\rho, K^{*})$, with $c$ quark as the spectator, which are CKM favored and thus have relatively large branching fractions.  In the evaluation,  the QCD factorization approach~\cite{Beneke:1999br,Beneke:2000ry} will be employed to calculate the hadronic matrix elements to the order of ${\cal O}(\alpha_s)$; moreover, not only the dominating longitudinal amplitude but also the power-suppressed transverse ones will be evaluated. In addition, the transition form factors as hadronic inputs will be estimated by using two different approaches: the Bauer-Stech-Wirbel~(BSW) model~\cite{Wirbel:1985ji,Bauer:1986bm,Bauer:1988fx} and the light-front quark model~(LFQM)~\cite{Jaus:1989au,Jaus:1991cy,Jaus:1999zv,Cheng:2003sm}.

This paper is organized as follows. The theoretical framework and calculations for $B_{c}$$\to$$ J/\psi V $ and $B_{c}^{*}$$\to$$\eta_{c}V$ decays are presented in section 2. Section 3 is devoted to the numerical
results and discussion. Finally, we give our summary in section 4.

  \section{Theoretical framework}
  \label{sec02}
  \subsection{The Amplitude in QCDF approach}
  \label{sec0201}
  The  effective Hamiltonian responsible for the $b\to c{\bar u}q$ $(q=d,s)$ induced $B_{c}$$\to$$ J/\psi V $ and $B_{c}^{*}$$\to$$\eta_{c}V$~($V=\rho, K^{*}$) decays can be written as~\cite{Buchalla:1995vs}
   \begin{equation}
  {\cal H}_{\rm eff}\ =\ \frac{G_{F}}{\sqrt{2}}\,
   \sum\limits_{q=d,s}\, V_{cb} V_{uq}^{\ast}\,
   \Big\{ C_{1}({\mu})\,Q_{1}({\mu})
         +C_{2}({\mu})\,Q_{2}({\mu}) \Big\}
   + {\rm h.c.}
   \label{hamilton},
   \end{equation}
  where $G_{F}$ is the Fermi coupling constant, and $ V_{cb} V_{uq}^{\ast}$ is the product of CKM matrix elements. The Wilson coefficients $C_{1,2}(\mu)$  summarize the physical contributions above the scale of ${\mu}$ and are calculable perturbatively~\cite{Buchalla:1995vs}; $Q_{1,2} $ are the corresponding local four-quark operators defined as
    \begin{eqnarray}
    Q_{1} &=&
  [ \bar{c}_{\alpha}{\gamma}_{\mu}(1-{\gamma}_{5})b_{\alpha} ]
  [ \bar{q}_{\beta} {\gamma}^{\mu}(1-{\gamma}_{5})u_{\beta} ]
    \label{q1}, \\
    Q_{2} &=&
  [ \bar{c}_{\alpha}{\gamma}_{\mu}(1-{\gamma}_{5})b_{\beta} ]
  [ \bar{q}_{\beta}{\gamma}^{\mu}(1-{\gamma}_{5})u_{\alpha} ]
    \label{q2},
    \end{eqnarray}
  where ${\alpha}$ and ${\beta}$ are color indices and the sum over repeated indices is understood.
 Then, the remaining work is to  calculate accurately the hadronic matrix elements of the local operators between initial and final states.

\begin{figure}[t]
\begin{center}
\includegraphics[scale=0.3]{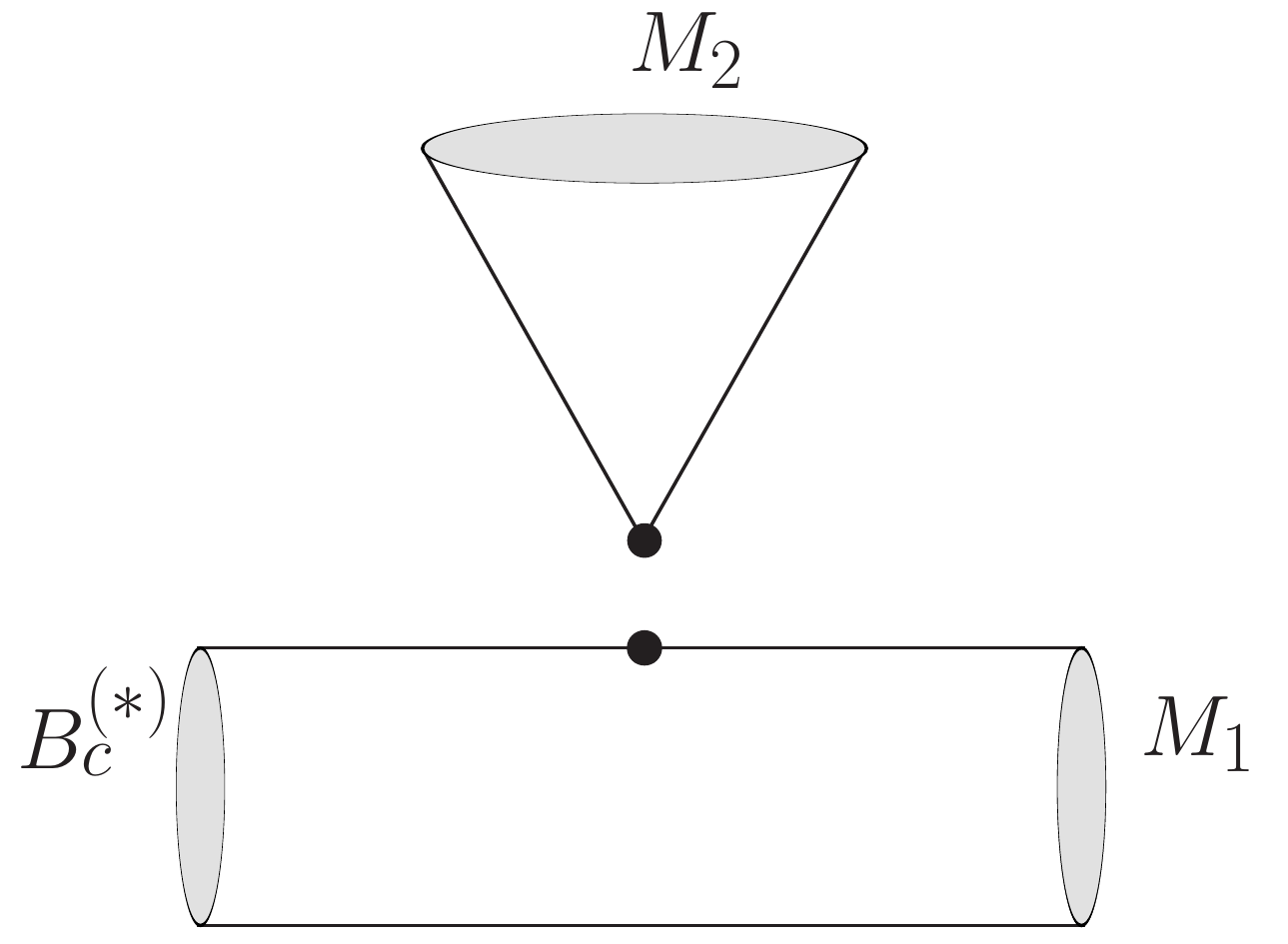}\quad
\caption{\label{fig:t} The leading-order contribution (tree diagram) .}
\end{center}
\end{figure}

\begin{figure}[t]
\begin{center}
\subfigure[]{\includegraphics[scale=0.3]{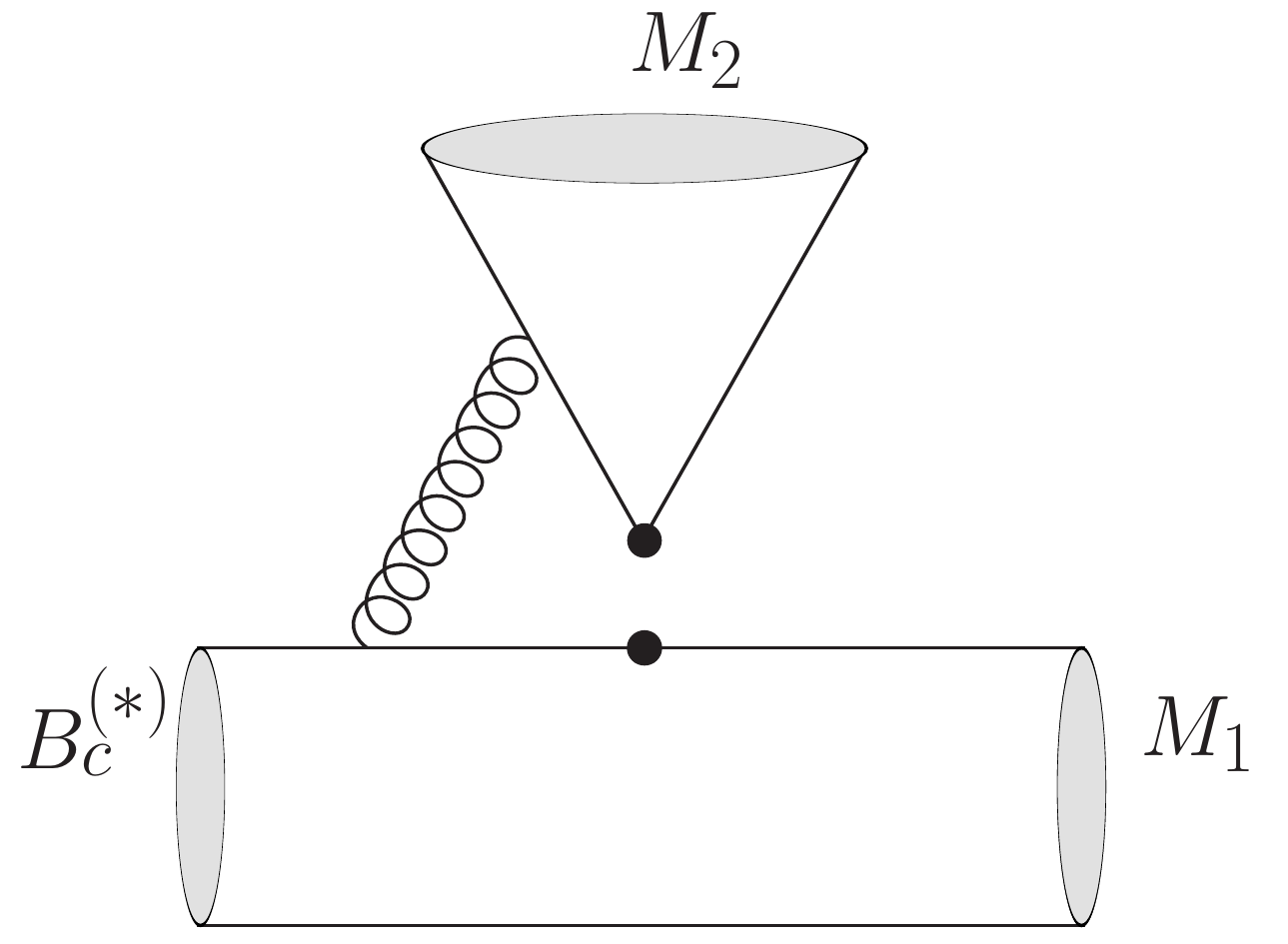}}\quad
\subfigure[]{\includegraphics[scale=0.3]{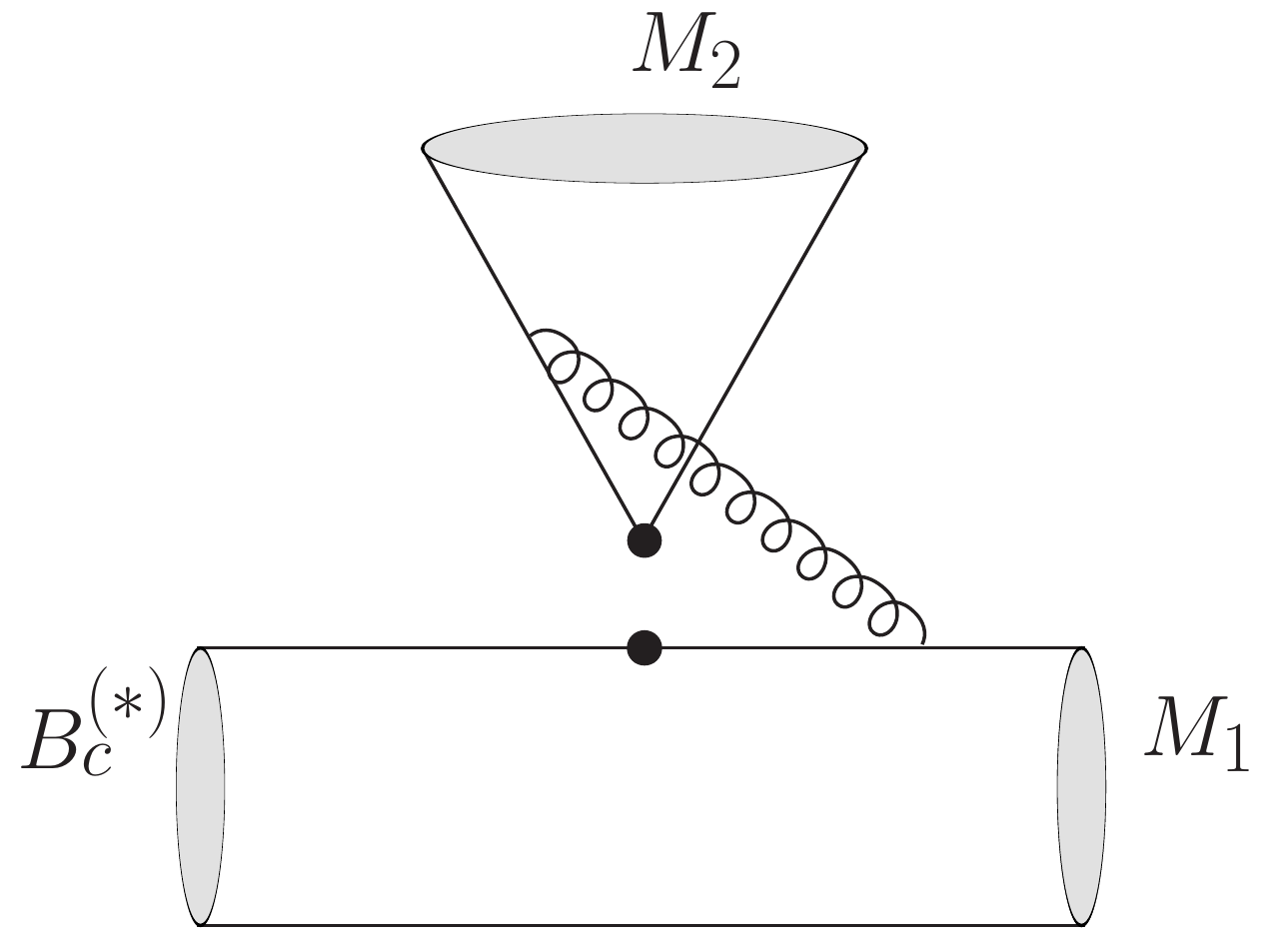}}\quad
\subfigure[]{\includegraphics[scale=0.3]{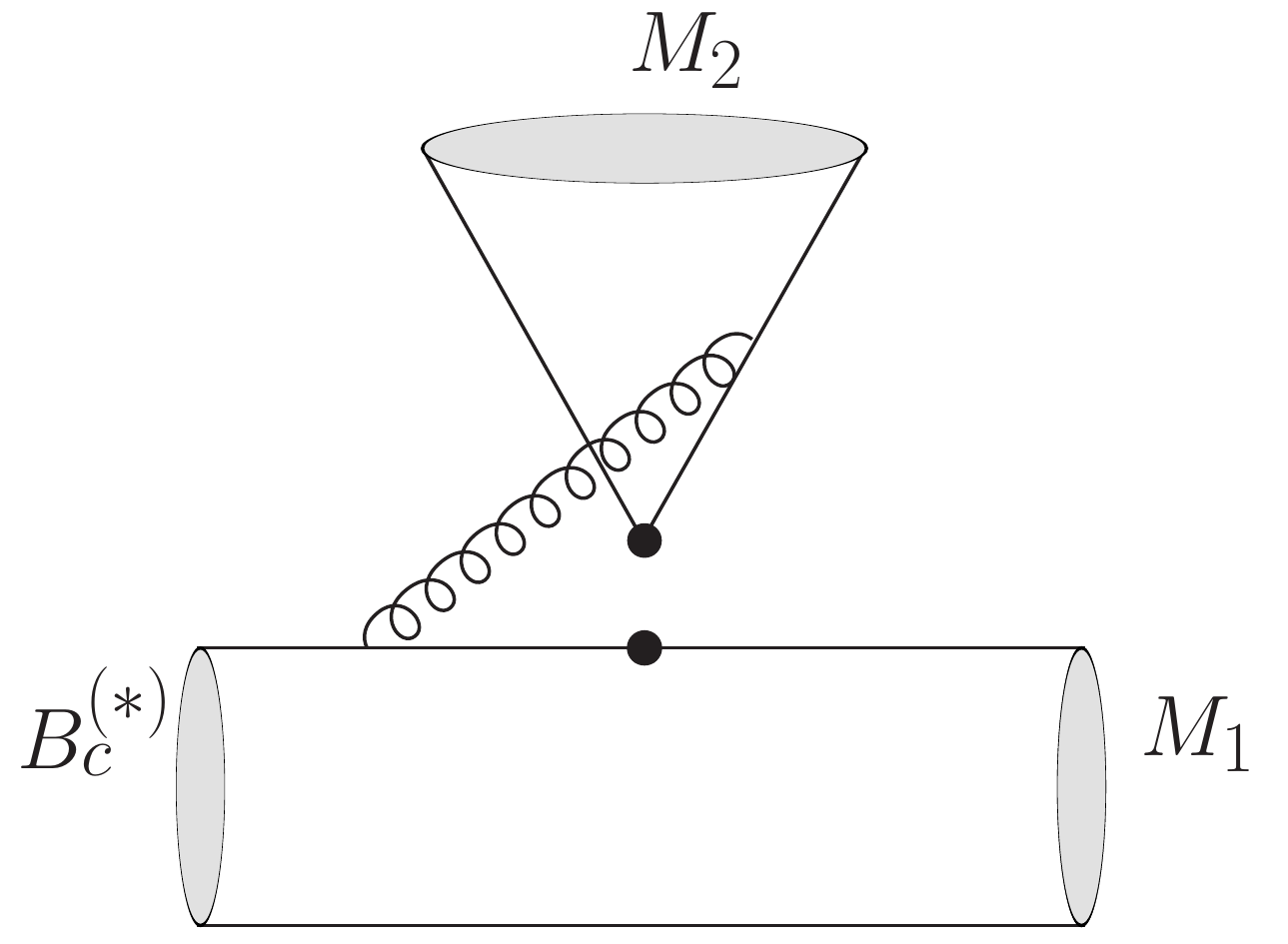}}\quad
\subfigure[]{\includegraphics[scale=0.3]{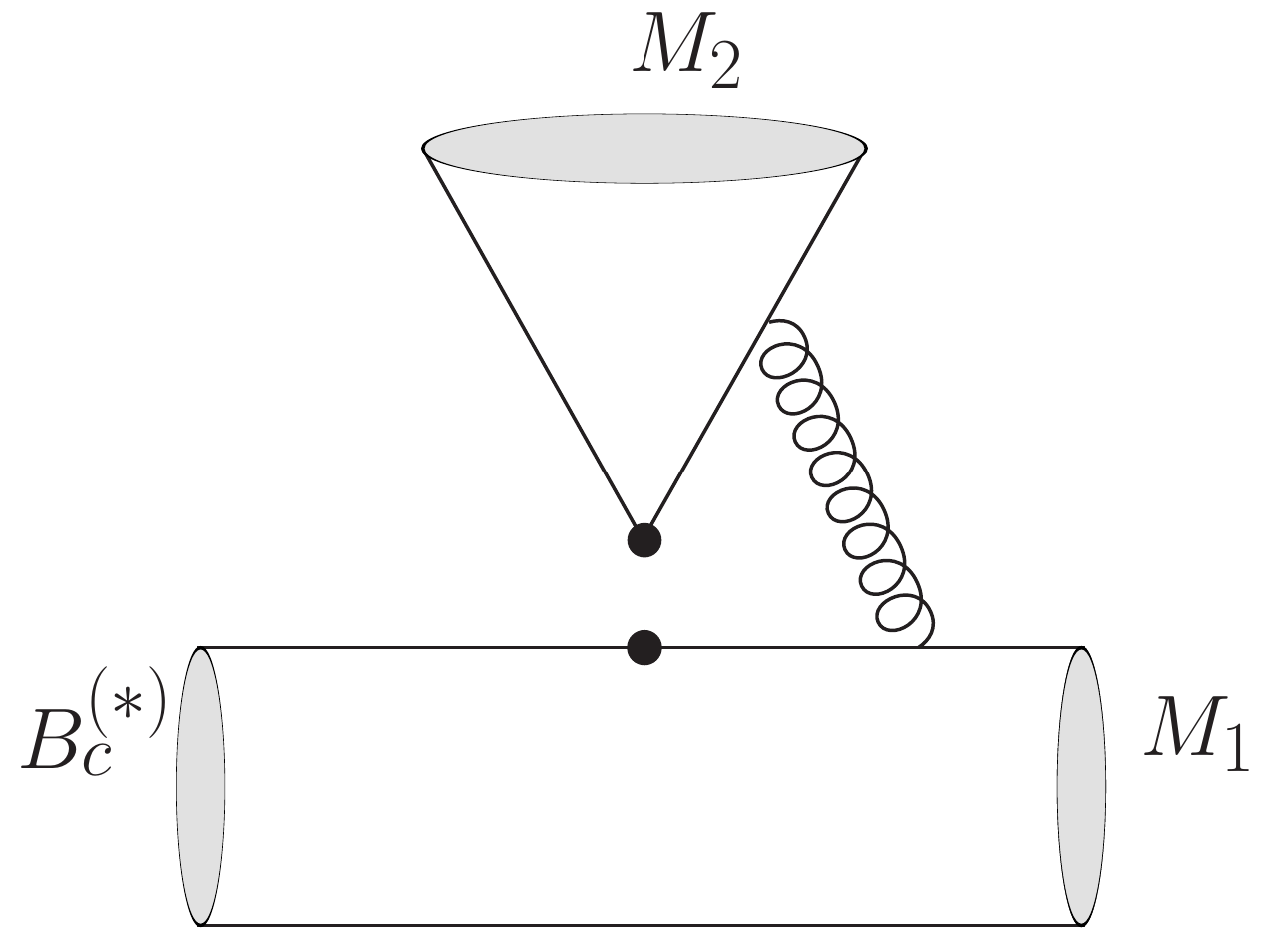}}
\caption{\label{fig:v} The vertex corrections at the order of $\alpha_s$.}
\end{center}
\end{figure}

  In order to take the QCD corrections into account, the QCDF approach is employed in this work.
 In this approach, the hadronic matrix element can be written as the convolution integrals of hard scattering
  kernel and universal light-cone distribution amplitude~(LCDA)~\cite{Beneke:1999br,Beneke:2000ry},
   \begin{equation}
  {\langle}M_{1}M_{2}{\vert}Q_{i}{\vert}{B_{c}^{(*)}}{\rangle} =
   \sum\limits_{j} F_{j}^{ {B_{c}^{(*)}}{\to}M_{1} }
  {\int}\,dx\, H_{ij}(x)\,{\Phi}_{M_{2}}(x)
   \label{hadronic},
   \end{equation}
  where the transition form factor $F_{j}^{ {B_{c}^{(*)}}{\to}M_{1} }$ and LCDA ${\Phi}_{M_{2}}(x)$ of the emitted meson are universal nonperturbative inputs; while, the hard scattering function, $H_{ij}(x)$, is calculable order by order through perturbative QCD theory. It is noted that such factorization formula is valid only when the the ``emission particle''~$M_2$ is light~\cite{Beneke:2000ry}.  
 For the case of  $B_{c}\to J/\psi V$ and $B_{c}^{*}\to\eta_{c}V$  decays, $M_{1}=J/\psi\,,\eta_{c}$ and $M_2=V$~(light vector meson); the hard scattering kernel $H_{ij}(x)$ receives the contributions from the tree and vertex diagrams shown by Figs.~\ref{fig:t} and \ref{fig:v},  while the penguin diagram doesn't exist in the $b\to c{\bar u}q$ $(q=d,s)$ transitions~(the flavors of final quarks are different with each other).
 
  Then, using the factorization formula, Eq.~(\ref{hadronic}), the helicity amplitudes can be written as
   \begin{eqnarray}
   \label{amp1}
  {\cal A}_{\lambda}({B_{c}}{\to}J/\psi V) &=&
  {\langle}J/\psi V{\vert}{\cal H}_{\rm eff}{\vert}{B_{c}}{\rangle} =
   \frac{G_{F}}{\sqrt{2}}\, V_{cb} V_{uq}^{\ast}\, a_{1}^{\lambda}\,
  H_{\lambda}\,,\\
   {\cal A}_{\lambda}({B_{c}^{*}}{\to}{\eta_{c} V}) &=&
  {\langle}{\eta_{c} V}{\vert}{\cal H}_{\rm eff}{\vert}{B_{c}^{*}}{\rangle} =
   \frac{G_{F}}{\sqrt{2}}\, V_{cb} V_{uq}^{\ast}\, a_{1}^{\lambda}\,
  H^{'}_{\lambda}\,,
   \label{amp2}
   \end{eqnarray}
  where $\lambda=0,\pm$ denotes the helicity of $V$ meson.

  In Eqs.~\eqref{amp1} and \eqref{amp2}, $H_{\lambda}$ and $H^{'}_{\lambda}$ are the product of matrix elements of the current operators,
  $H_{\lambda}\equiv {\langle }V|\bar{q}{\gamma}^{\mu}(1-{\gamma}_{5})u|0\rangle
  {\langle} J/\psi|\bar{c}{\gamma}_{\mu}(1-{\gamma}_{5})b| {B_{c}}{\rangle}$  and
  $H^{'}_{\lambda}\equiv {\langle }V|\bar{q}{\gamma}^{\mu}(1-{\gamma}_{5})u|0\rangle
  {\langle} \eta_{c}|\bar{c}{\gamma}_{\mu}(1-{\gamma}_{5})b|{B^{*}_{c}}{\rangle}$.
 The current matrix elements can be factorized by the decay constant and form factors, which are defined as 
   \begin{eqnarray}
    {\langle}V({\varepsilon_2},p_2){\vert}{\bar q}\gamma^{\mu}u{\vert}0{\rangle}
  &=&-i
   f_{V}\,m_{V}\,{\varepsilon}_{2}^{\ast\mu}
   \label{cme02},
   \end{eqnarray}
for the vector meson,
    \begin{eqnarray}
\langle J/\psi(\varepsilon_1,p_1)|\bar{c}\gamma_{\mu} b|{B_c}(p)\rangle &=&
\frac{2iV(q^2)}{m_{B_c}+m_{J/\psi}}\epsilon_{\mu\nu\rho\sigma}\varepsilon^{*\nu}_1p^{\rho}p_{1}^{\sigma},\\
\langle J/\psi(\varepsilon_1,p_1)|\bar{c}\gamma_{\mu}\gamma_5 b|{B_c}( p)\rangle &=&
2m_{J/\psi}A_0(q^2)\frac{\varepsilon_1^* \cdot q}{q^2}q_{\mu}
+(m_{J/\psi}+m_{B_c})A_1(q^2)\left(\varepsilon^*_{1\mu}-\frac{\varepsilon^*_1\cdot q}{q^2}q_{\mu}\right)\nonumber \\
&&-A_2(q^2)\frac{\varepsilon_1^* \cdot q}{m_{J/\psi}+m_{B_c}} \left[(p_1+p)_{\mu}-\frac{m^2_{B_c}-m^2_{J/\psi}} {q^2} q_{\mu}\right],
\end{eqnarray}
 for the ${B_c}\to J/\psi$ transition, and
\begin{eqnarray}
\langle \eta_c(p_{1})|\bar{c}\gamma_{\mu} b|B^*_{c}(\varepsilon, p)\rangle &=&
\frac{2iV(q^2)}{m_{B_c^*}+m_{\eta_c}}\epsilon_{\mu\nu\rho\sigma}\varepsilon^{\nu}p^{\rho}p_{1}^{\sigma},\\
\langle \eta_c(p_{1})|\bar{c}\gamma_{\mu}\gamma_5 b|B^*_{c}(\varepsilon, p)\rangle &=&
2m_{B_c^*}A_0(q^2)\frac{\varepsilon \cdot q}{q^2}q_{\mu}
+(m_{\eta_c}+m_{B_c^*})A_1(q^2)\left(\varepsilon_{\mu}-\frac{\varepsilon\cdot q}{q^2}q_{\mu}\right)\nonumber \\
&&+A_2(q^2)\frac{\varepsilon \cdot q}{m_{\eta_c}+m_{B^*_{c}}} \left[(p+p_{1})_{\mu}-\frac{m^2_{B_c^*}-m^2_{\eta_c}} {q^2} q_{\mu}\right],
\end{eqnarray}
for the $B^*_{c}\to \eta_c$ transition.
Here, the sign convention $\epsilon_{0123}=-1$ is used.
Then, we can finally obtain the expressions of $H_{\lambda}$ and $H^{'}_{\lambda}$ by contracting the current matrix elements. Explicitly, they are written as
\begin{eqnarray}
\label{eq:H0B}
H_0&=&\frac{i\,f_V}{2\,m_{J/\psi}}\left[(m_{B_{c}}^2-m_{J/\psi}^2-m_V^2)(m_{B_{c}}+m_{J/\psi})A_1^{{B_{c}}\to {J/\psi}}(m_V^2)-\frac{4\,m_{B_{c}}^2\,p_c^2}{m_{B_{c}}+m_{J/\psi}}A_2^{{B_{c}}\to {J/\psi}}(m_V^2)\right],\nonumber\\
\\
\label{eq:HmpB}
H_{\mp}&=&i\,f_V\,m_V\left[(m_{B_{c}}+m_{J/\psi})A_1^{{B_{c}}\to {J/\psi}}(m_V^2)\pm \frac{2\,m_{B_{c}}\,p_c}{m_{B_{c}}
+m_{J/\psi}}V^{{B_{c}}\to {J/\psi}}(m_V^2)\right],
\end{eqnarray}
for ${B_c}\to J/\psi V$ decays, and
\begin{eqnarray}
\label{eq:H0Bstar}
H_0^{\prime}&=&\frac{i\,f_V}{2\,m_{B_{c}^*}}\left[(m_{B_{c}^*}^{2}-m_{\eta_{c}}^2+m_V^2)(m_{B_{c}^*}+m_{\eta_{c}})A_1^{B_{c}^*\to \eta_{c}}(m_V^2)+\frac{4\,m_{B_{c}^*}^2\,p_c^{\prime 2}}{m_{B_{c}^*}+m_{\eta_{c}}}A_2^{B_{c}^*\to \eta_{c}}(m_V^2)\right]\,,\nonumber\\
\\
\label{eq:HmpBstar}
H_{\mp}^{\prime}&=&-i\,f_V\,m_V\left[(m_{B_{c}^*}+m_{\eta_{c}})A_1^{B_{c}^*\to \eta_{c}}(m_V^2)\pm \frac{2\,m_{B_{c}^*}\,p_c^{\prime}}{m_{B_{c}^*}+m_{\eta_{c}}}V^{B_{c}^*\to \eta_{c}}(m_V^2)\right],
\end{eqnarray}
for $B^*_{c}\to {\eta_c}V$ decays, where
$p_c=\sqrt{[m_{B_c}^2-(m_{J/\psi}+m_V)^2][m_{B_c}^2-(m_{J/\psi}-m_V)^2]}/(2m_{B_c})$
and $p_c^{\prime}$ is obtained from $p_c$ by replacing $m_{B_c}\to m_{B_c^*}$ and $m_{J/\psi}\to m_{\eta_c}$.

  The $a_{1}^{\lambda}$  in  Eqs.~\eqref{amp1} and \eqref{amp2} are effective coefficients and written as
  \begin{equation}
   a_{1}^{\lambda}
    = C_{1}^{\rm NLO}+\frac{1}{N_{c}}\,C_{2}^{\rm NLO}
    + \frac{{\alpha}_{s}}{4{\pi}}\, \frac{C_{F}}{N_{c}}\,
      C_{2}^{\rm LO}\, V^{\lambda}_1\,,
   \label{a1}
  \end{equation}
 where the first two terms are the contributions of tree diagrams, and $V^{\lambda}_1$ is the vertex function obtained by calculating  QCD vertex diagrams~\cite{Beneke:2000ry}. After calculation, it can be found that only the leading-twist LCDA of emitted vector meson contributes to  $V_1^0$ and twist-3 ones contribute to  $V_1^\mp$. In the previous works, for instance Ref.~\cite{Beneke:2000ry}, the transverse contributions are usually neglected because they are power suppressed. In this paper,  the full contributions, $a_{1}^{0,\mp}$, are  taken into account, and we will show that the transverse amplitudes provide nontrivial corrections to the branching fractions of $B_{c}$$\to$$ J/\psi V $ and $B_{c}^{*}$$\to$$\eta_{c}V$ decays.

 The explicit expressions of the vertex corrections $V_1^{\lambda}$ are
  \begin{eqnarray}\label{eq:v10}
  V_1^0 ={\int}_{0}^{1}du\, {\Phi}_{V}(u)
  \left[ 3\,{\log} \Big( \frac{ m_{b}^{2} }{ {\mu}^{2} } \Big)+ 3\,{\log} \Big( \frac{ m_{c}^{2} }{ {\mu}^{2} } \Big)- 18 +g_0(u)\right]\,,\\
  \label{eq:vmp}
  V_1^{\mp} ={\int}_{0}^{1}du\, \,{\Phi}_{\mp}(u)
  \left[ 3\,{\log} \Big( \frac{ m_{b}^{2} }{ {\mu}^{2} } \Big)+ 3\,{\log} \Big( \frac{ m_{c}^{2} }{ {\mu}^{2} } \Big)- 18 +g_{\mp}(u)\right]\,,
  \end{eqnarray}
where  $u$ is the longitudinal momentum fraction variable of quark in the emission vector meson; ${\Phi}_{V}(u)$ is the leading-twist LCDA, and conventionally expanded in Gegenbauer polynomials~\cite{Ball:1998je,Ball:2006wn},
\begin{eqnarray}
\label{eq:twist-2}
 {\Phi}_{V}(u)=6u\bar{u}\left[ 1+\sum_{n=1}^{\infty}\, a_{n}^{V}(\mu)\,C_{n}^{3/2}(2u-1)\right]\,
\end{eqnarray}
with $\bar{u}\equiv(1-u)$; while ${\Phi}_{\mp}(u)$ (${\Phi}_{-}(u)=\phi_b$ and ${\Phi}_{+}(u)=\phi_a$) are the twist-3 LCDAs, and given by
\begin{eqnarray}
\label{eq:twist-3}
\phi_a(u)=\int_{u}^{1}dv\frac{{\Phi}_{V}(v)}{v}\,,    \qquad
\phi_b(u)=\int_{0}^{u}dv\frac{{\Phi}_{V}(v)}{\bar{v}}\,.
\end{eqnarray}
The loop functions $g_{0,\mp}(u)$ in Eqs.~(\ref{eq:v10}) and (\ref{eq:vmp}) read
  \begin{eqnarray}
 g_0(u)&=&
      \frac{c_{a}}{1-c_{a}}\,{\log}(c_{a})-\frac{4\,c_{b}}{1-c_{b}}\,{\log}(c_{b})
      +\frac{c_{d}}{1-c_{d}}\,{\log}(c_{d})-\frac{4\,c_{c}}{1-c_{c}}\,{\log}(c_{c})
      \nonumber \\
      &&+f(c_{a})-f(c_{b})-f(c_{c})+f(c_{d})
      +2\, {\log}(r_{c}^{2}) \big[ {\log}(c_{a}) -{\log}(c_{b}) \big]  \nonumber \\
      &&+r_{c}\, \Big[ \frac{c_{a}}{(1-c_{a})^{2}}\,{\log}(c_{a})
    + \frac{1}{1-c_{a}} \Big]
      +r_{c}^{-1}\,\Big[ \frac{c_{d}}{(1-c_{d})^{2}}\,{\log}(c_{d})
    + \frac{1}{1-c_{d}} \Big]\,,
   \end{eqnarray}
     \begin{eqnarray}
       g_{\mp}(u)&=&
      \frac{1+c_{a}}{1-c_{a}}\,{\log}(c_{a})-\frac{4\,c_{b}}{1-c_{b}}\,{\log}(c_{b})
      +\frac{1+c_{d}}{1-c_{d}}\,{\log}(c_{d})-\frac{4\,c_{c}}{1-c_{c}}\,{\log}(c_{c})
      \nonumber \\
      &&+f(c_{a})-f(c_{b})-f(c_{c})+f(c_{d})
      +2\, {\log}(r_{c}^{2}) \big[ {\log}(c_{a}) -{\log}(c_{b}) \big]\nonumber \\
     && +k_{\mp}r_{c}\, \Big[ \frac{2c_{a}-1}{(1-c_{a})^{2}}\,{\log}(c_{a})
    + \frac{1}{1-c_{a}} \Big]
     +k_{\mp}r_{c}^{-1}\,\Big[ \frac{2c_{d}-1}{(1-c_{d})^{2}}\,{\log}(c_{d})
    + \frac{1}{1-c_{d}} \Big]\,,
   \end{eqnarray}
 where $r_{c} = m_{c}/m_{b}$, $c_{a} = u\,(1-r_{c}^{2})$, $c_{b} = \bar{u}\,(1-r_{c}^{2})$,
  $c_{c} = -c_{a}/r_{c}^{2}$, $c_{d} = -c_{b}/r_{c}^{2}$, $f(c)=2{\rm Li}_2(\frac{c-1}{c})-\log^2(c)-\frac{2c}{1-c}\log(c)$, and
 $k_{\mp}$${\equiv}$$-\tilde{H}_{\mp}/H_{\mp}$ with
 $\tilde{H}_{\mp}=H_{\mp}$($A_{1}\to{-A_{1}}$). It can be easily checked that the results for $B\to V_1V_2$ decays ($V$ denotes light vector meson), for instance the Eqs. (A.7) and (A.8) in Ref. \cite{Beneke:2006hg}, can be recovered  from Eqs. (\ref{eq:v10}) and (\ref{eq:vmp}) by taking the limit $m_{c}\to 0$.

  Using the decay amplitudes given above, the branching fractions of
   $B_{c}\to J/\psi V$ and $B_{c}^{*}\to\eta_{c}V$ decays in the center-of-mass frame of $B_{c}^{(*)}$ are defined as
  \begin{eqnarray}
  {\cal B}(B_{c}\to J/\psi V) &=&\frac{1}{8\pi}\frac{p_{c}}{m^2_{B_{c}}\Gamma_{tot}({B_{c}})}
  \sum_{\lambda}|{\cal A}_{\lambda}({B_{c}}\to J/\psi V)|^2\,\\
  {\cal B}({B_c^*}\to \eta_{c} V) &=&\frac{1}{24\pi}\frac{p_{c}^{\prime}}{m^2_{{B_c^*}}\Gamma_{tot}({B_c^*})}
  \sum_{\lambda}|{\cal A}_{\lambda}({B_c^*}\to \eta_{c} V)|^2\,
  \end{eqnarray}
  where $\Gamma_{tot}(B_{c})$ and $\Gamma_{tot}(B_{c}^{*})$ are the total decay width of
   $ {B_{c}}$ and ${B_{c}^{*}}$ meson, respectively.
   Along with the branching fraction, the polarization fractions are defined as
   \begin{eqnarray}
   f_{L,{\parallel},{\perp}} =
   \frac{ {\vert}{\cal A}_{0,{\parallel},{\perp}}{\vert}^{2} }
        { {\vert}{\cal A}_{0}{\vert}^{2}
         +{\vert}{\cal A}_{\parallel}{\vert}^{2}
         +{\vert}{\cal A}_{\perp}{\vert}^{2} }
   \end{eqnarray}
   where ${\cal A}_{\parallel}$ and ${\cal A}_{\bot}$ are parallel and perpendicular amplitudes,
   and can be easily obtained through
${\cal A}_{\parallel,\bot} = ({\cal A}_{-}{\pm}{\cal A}_{+})/\sqrt{2}$.

  \subsection{Form factors in the BSW model and LFQM}
  \label{sec0203}

 The form factors $A_1$, $A_2$ and $V$ as nonperturbative  inputs
  are essential in evaluating $H_{\lambda}$ and $H^{\prime}_{\lambda}$ .
  However, because of lacking the experimental information, we employ two fully developed approaches, Bauer-Stech-Wirbel~(BSW) model~\cite{Wirbel:1985ji,Bauer:1988fx} and light-front quark model~(LFQM)~\cite{Jaus:1989au,Jaus:1991cy,Jaus:1999zv,Cheng:2003sm}, to estimate the values of form factors.

 In the  BSW model, the form factors could be expressed as the overlap integrals of the initial and final meson wave functions. The form factors, $A_{0,1}$ and $V$ at $q^{2}$$=$$0$ are written as~\cite{Wirbel:1985ji,Bauer:1986bm,Bauer:1988fx}
  \begin{eqnarray}
  A_{0}^{{B_{c}}{\to}J/\psi}(0) &=&
  {\int}d\vec{k}_{\perp} {\int}_{0}^{1}dx\,
   \Big\{ {\Phi}_{J/\psi}(\vec{k}_{\perp},x,1,0)\,
  {\sigma}_{z} {\Phi}_{B_{c}}(\vec{k}_{\perp},x,0,0)\Big\}
  \label{form-a0},\\
  A_{1}^{ {B_{c}}{\to} J/\psi}(0) &=&
  \frac{ m_{b}+m_{c} }{ m_{B_{c}}+m_{J/\psi }}
  I^{ {B_{c}}{\to} J/\psi}\,,
  \label{form-a1} \\
  V^{ {B_{c}}{\to} J/\psi }(0)  &=&
  \frac{ m_{b}-m_{c} }{ m_{B_{c}}-m_{J/\psi } }
  I^{ {B_{c}}{\to} J/\psi}
  \label{form-v},
  \end{eqnarray}
with
  \begin{equation}
  I^{{B_{c}}{\to} J/\psi} = \sqrt{2}
  {\int}d\vec{k}_{\perp} {\int}_{0}^{1} \frac{dx}{x}\,
   \Big\{ {\Phi}_{J/\psi}(\vec{k}_{\perp},x,1,-1) \,
  i{\sigma}_{y}\,{\Phi}_{B_c}(\vec{k}_{\perp},x,0,0)\Big\}
  \label{form-ii},
  \end{equation}
  where ${\sigma}_{y,z}$ is a Pauli matrix acting on
  the spin indices of the decaying bottom quark;
  $x$ and $\vec{k}_{\perp}$ denote the fraction of
  the longitudinal momentum and the transverse momentum
  carried by the nonspectator quark, respectively; $m_c=1.7\,{\rm GeV}$ and $m_b= 4.9\,{\rm GeV}$ are the constituent quark masses. The form factor $A_{2}(0)$ can be obtained through the relation
   \begin{equation}
    A_{2}^{ {B_{c}}{\to} J/\psi }(0) =
  \frac{ m_{B_{c}}+m_{J/\psi }}{ m_{B_{c}}-m_{J/\psi }}A_{1}^{ {B_{c}}{\to} J/\psi }(0)
  -\frac{2m_{J/\psi}}{ m_{B_{c}}-m_{J/\psi }}A_{0}^{ {B_{c}}{\to} J/\psi }(0)\,.
    \end{equation}
    The form factors for $B_c^*\to\eta_c$ transition can be obtained through the replacement $ {\Phi}_{B_c}(\vec{k}_{\perp},x,0,0)\to {\Phi}_{B_c^*}(\vec{k}_{\perp},x,1,s_z)$ ~($s_z=0$ and $-1$ for $A_{0}^{{B_{c}^*}{\to}\eta_c}$ and $I^{{B_{c}^*}{\to}\eta_c }$, respectively), ${\Phi}_{J/\psi}(\vec{k}_{\perp},x,1,-1) \to  {\Phi}_{\eta_c}(\vec{k}_{\perp},x,0,0)$, $m_{B_c}\to m_{B_c^*}$ and $m_{J/\psi}\to m_{\eta_c}$.
  In the evaluation, the orbital part of the meson wavefunction~\cite{Wirbel:1985ji,Bauer:1988fx},
    \begin{equation}
  \label{wave}
  {\phi_{M}}({\vec k}_{{\perp}},x)=N_{M}\sqrt{x(1-x)}\,{\rm exp}\left(-{\vec k}^{2}_{{\perp}}/2{\omega^{2}}\right)
  \,{\rm exp}\left[-\frac{M^2}{2\omega^{2}}\left(x-\frac{1}{2}-\frac{m_{q_1}^{2}-m_{q_2}^{2}}{2M^{2}}\right)^2\right]\,,
  \end{equation}
is used, where $N_{M}$ is the normalization factor, and the parameter $\omega$ determines the average transverse momentum, $\langle{\vec k}_{{\perp}}^2\rangle= \omega^2$.

 In order to ensure the reliability of theoretical results, we will use the LFQM in addition to the BSW model to reevaluate the form factors  and further compare their results. The LFQM has been fully developed in Refs.~\cite{Jaus:1989au,Jaus:1991cy,Jaus:1999zv,Cheng:2003sm}.  The form factors  used in this paper are related to the convention used in the LFQM through
  \begin{eqnarray}
 \label{lf:ff2}
 V(q^2)&=&(M_{1}+M_{2})g(q^2)\,,\\
 A_{0}(q^2)&=&-\frac{1}{2M_{2}}[f(q^2)+(M_{1}^{2}-M_{2}^{2})a_{+}(q^{2})+q^{2}a_{-}(q^2)]\,,\\
 A_{1}(q^{2})&=&-\frac{f(q^2)}{M_{1}+M_{2}}\,,\\
  A_{2}(q^{2})&=&(M_{1}+M_{2})a_{+}(q^{2})\,,
 \end{eqnarray}
 where $M_{1}$ and  $M_{2}$ are the masses of the initial and final mesons, respectively; $g(q^2)$, $f(q^2)$ and $a_{+\,,-}(q^2)$ are another set of independent form factors defined by
  \begin{eqnarray}
  \label{lf:ff1}
 &&\langle V(p_{2},\varepsilon)|\bar{q}_{2}\gamma^{\mu}(1-\gamma_{5})q_{1}|P(p_{1})\rangle\nonumber\\
 &=&i\epsilon_{\mu\nu\rho\sigma}\varepsilon^{*\nu}P^{\rho}q^{\sigma}g(q^2)+
  \varepsilon^{*\mu}f(q^2)+P^{\mu}(\varepsilon^{*}\cdot P)a_{+}(q^{2})
  +q^{\mu}(\varepsilon^{*}\cdot P)a_{-}(q^{2})
  \end{eqnarray}
 where $P=p_{1}+p_{2}$ and $q=p_{1}-p_{2}$.  At the level of one-loop approximation, the explicit expression of the form factors are given by~\cite{Choi:1999tk}
 \begin{eqnarray}
 \label{lf:ff2}
 g(q^{2})&=&\int^{1}_{0}dx \int d^{2}\textbf{k}_{\perp}
 \frac{x\phi_{2}^{*}(x,\textbf{k}_{\perp}^{'})\phi_{1}(x,\textbf{k}_{\perp})}{\sqrt{{\cal A}_{2}^2+\textbf{k}_{\perp}^{'2}}
 \sqrt{{\cal A}_{1}^2+\textbf{k}_{\perp}^{2}}}\nonumber\\
 &&\times\bigg\{{\cal A}_{1}-\frac{m_{q_1}-m_{q_2}}{\textbf{q}_{\perp}^{2}}\textbf{k}_{\perp}\cdot \textbf{q}_{\perp}+
 \frac{2}{M_{20}+m_{q_2}+m_{\bar q}}\big[\textbf{k}_{\perp}^{2}-\frac{(\textbf{k}_{\perp}\cdot \textbf{q}_{\perp})^2}{\textbf{q}_{\perp}^2}\big]
 \bigg\},\\
    \label{lf:ff3}
 a_{+}(q^{2})&=&\int^{1}_{0}dx \int d^{2}\textbf{k}_{\perp}
 \frac{x\phi_{2}^{*}(x,\textbf{k}_{\perp}^{'})\phi_{1}(x,\textbf{k}_{\perp})}{\sqrt{{\cal A}_{2}^2+\textbf{k}_{\perp}^{'2}}
 \sqrt{{\cal A}_{1}^2+\textbf{k}_{\perp}^{2}}}\nonumber\\
 &&\times\bigg\{(1-2x){\cal A}_{1}-\frac{\textbf{k}_{\perp}\cdot \textbf{q}_{\perp}}{x\textbf{q}_{\perp}^{2}}
 [(1-2x){\cal A}_{1}-{\cal A}_{2}]
-2\frac{(1-\textbf{k}_{\perp}\cdot \textbf{q}_{\perp}/{x\textbf{q}_{\perp}^{2}})}
 {M_{20}+m_{q_2}+m_{\bar q}}(\textbf{k}_{\perp}^{'}\cdot {\textbf{k}_{\perp}} +{\cal A}_{1}{\cal B}_{2})
\bigg\},\nonumber\\
\\
\label{lf:ff4}
f(q^2)&=&(M_{2}^{2}-M_{1}^{2}+q^{2})a_{+}(q^2)\nonumber\\
&&-2M_{2}\int_{0}^{1}dx\int d^{2}\textbf{k}_{\perp}
 \frac{\phi_{2}^{*}(x,\textbf{k}_{\perp}^{'})\phi_{1}(x,\textbf{k}_{\perp})}{\sqrt{{\cal A}_{2}^2+\textbf{k}_{\perp}^{'2}}
 \sqrt{{\cal A}_{1}^2+\textbf{k}_{\perp}^{2}}}\nonumber\\
 &&\times\bigg\{2x(1-x){{\cal A}_{1}}M_{20}+\frac{(1-2x)M_{20}+m_{q_2}-m_{\bar q}}{M_{20}+m_{q_2}+m_{\bar q}}
[\textbf{k}_{\perp}^{'}\cdot {\textbf{k}_{\perp}} +{\cal A}_{1}{\cal B}_{2}]
\bigg\}\,,
 \end{eqnarray}
 where ${\bf k}_{\perp}^{'}={\bf k}_{\perp}-x{\bf q}_{\perp}$; ${\cal A}_{n}=xm_{q_n}+(1-x)m_{\bar q}$~(n=1,2) and ${\cal B}_{n}=-xm_{q_n}+(1-x)m_{\bar q}$ with $q_{1,2}=b,c$ and $\bar q=\bar c$ for $\bar{B}_c\to J/\Psi$ transition; $M_{n0}$ with $n=1$ and $2$ denote the invariant mass of the initial and final states, respectively, and are written as
  \begin{eqnarray}
 M_{n0}^{2}=\frac{\textbf{k}_{\perp}^{2}+m_{q_{n}}^{2}}{x}+\frac{\textbf{k}_{\perp}^{2}+m_{\bar q}^{2}}{1-x}\,.
 \end{eqnarray}

In the Eqs.~\eqref{lf:ff2}, \eqref{lf:ff3} and \eqref{lf:ff4}, $\phi_{1,2}(x,\textbf{k}_{\perp})$ are the radial wavefunctions of the initial and final states, respectively. In the LFQM and this paper, the general form of Gaussian-type wavefunction is used. It is given by
 \begin{eqnarray}
 \label{eq:wavelf1}
 && \phi_{n}(x, {\bf k}_{\perp})= N\frac{\pi ^{1/3}}{\beta^{3/2}}
 \sqrt{\frac{\partial k_z}{\partial x}}
 \exp\left[\frac{-(k_{z}^{2}+{\bf k}_{\perp}^{2})}{2\beta_{n}^{2}}\right]\,,
 \end{eqnarray}
where $\beta_{n}$ is the mass scale parameter and can be obtained by fitting to data; $k_{z}=(x-1/2)M_{n0}+\frac{m_{\bar q}^{2}-m_{n}^{2}}{2M_{n0}}$ is the longitudinal component;  $\partial k_z/\partial x$ is the Jacobian factor of  variable transformation $(x,{\bf k}_{\perp})\to \vec{ k}$; $N$ is the normalization constant determined by
  \begin{eqnarray}
 \label{eq:wavelf2}
 && \int_{0}^{1}{\rm d x}\, {\rm d}^{2}{\bf k}_{\perp}|\phi_{n}(x, {\bf k}_{\perp})|^{2}=1\,.
 \end{eqnarray}

  \section{Numerical results and discussion}
     \label{sec03}
   \begin{table}[t]
   \caption{Numerical values of input parameters.}
   \label{tab:input}
  \begin{center}\setlength{\tabcolsep}{5pt}
   \begin{tabular}{llll}\hline\hline
Wolfenstein parameters &
    ${\lambda}$$=$$0.22543^{+0.00042}_{-0.00031}$,  $A$$=$$0.8227^{+0.0066}_{-0.0136}$\,;\cite{Patrignani:2016xqp} \\ \hline
masses of charm and bottom quarks&
    $m_{c}$ $=$ $1.67{\pm}0.07$ GeV, $m_{b}$ $=$ $4.78{\pm}0.06$ GeV\,;\cite{Patrignani:2016xqp}\\ \hline
decay constants &
   $f_{\rho}$$=$$216{\pm}3$ MeV\,, $f_{K^{\ast}}$$=$$220{\pm}5$ MeV\,;\cite{Ball:2007rt} \\ \hline
Gegenbauer moments at ${\mu}$$=$$2$ GeV &
$a_{1}^{\rho}$$=$$0$, $a_{2}^{\rho}$$=$$0.10$; $a_{1}^{K^{\ast}}$$=$$0.02$, $a_{2}^{K^{\ast}}$$=$$0.08$. \cite{Ball:2007rt} \\\hline\hline
 \end{tabular}
\end{center}
  \end{table}

In our numerical calculation,  the values of input parameters, including the CKM Wolfenstein parameters,
 masses of $b$ and $c$ quarks, the decay constants and Gegenbauer moments of distribution amplitudes
 in Eq.~(\ref{eq:twist-2}), are collected in Table \ref{tab:input}. For the other well-known inputs, such as the masses of mesons, the Fermi coupling constant $G_F$ and so on, we take their central values given by PDG~\cite{Patrignani:2016xqp}.

   \begin{table}[t]
   \caption{The theoretical predictions for $\Gamma(B^*_{c}\to B_{c}\gamma)$~\cite{Barik:1994vd,Eichten:1994gt,Kiselev:1994rc,Fulcher:1998ka,Ebert:2002pp,Godfrey:2004ya,Jena:2002is,Lahde:2002wj,Ciftci:2001kt} in units of eV.}
   \label{tab:rBcstar}
  \begin{center}\setlength{\tabcolsep}{5pt}
   \begin{tabular}{cccccccccc}\hline\hline
    Refs.   &\cite{Barik:1994vd} & \cite{Eichten:1994gt}&\cite{Kiselev:1994rc}&\cite{Fulcher:1998ka}&\cite{Ebert:2002pp}&\cite{Godfrey:2004ya} &\cite{Jena:2002is}&\cite{Lahde:2002wj}&\cite{Ciftci:2001kt}\\ \hline
$\Gamma(B^*_{c}\to B_{c}\gamma)$&$20$&$135$&$60$&$59$&$33$&$80$&$30$&$34$&$32$ \\\hline\hline
 \end{tabular}
\end{center}
  \end{table}

In order to evaluate the branching fractions of $B^*_{c}$ weak decays, the total decay width~(or lifetime) $\Gamma_{\rm{tot}}(B^*_{c})$ is essential. However,  there is no available experimental or theoretical information until now.  In this paper, due to the known fact that the radiative  process  $B^*_{c}\to B_{c}\gamma$ dominates the decays of $B^*_{c}$ meson, the approximation $\Gamma_{\rm{tot}}(B^*_{c})\simeq \Gamma(B^*_{c}\to B_{c}\gamma)$ is taken. The theoretical predictions for $\Gamma(B_{c}^*\to B_{c}\gamma)$
  have been given in some theoretical models~\cite{Barik:1994vd,Eichten:1994gt,Kiselev:1994rc,Fulcher:1998ka,Ebert:2002pp,Godfrey:2004ya,Jena:2002is,Lahde:2002wj,Ciftci:2001kt}; the numerical results are summarized in Table~\ref{tab:rBcstar} .
 Unfortunately, because the photon is not hard enough, these estimations suffer from large uncertainties. 
 From Table~\ref{tab:rBcstar}, it can be seen that most of the theoretical predictions are in the range $[20,80]$ eV except for $135$ eV given by Ref.~\cite{Eichten:1994gt}. Therefore, to give a quantitative estimation, we take a conservative choice, $\Gamma_{\rm{tot}}(B^*_{c})\simeq\Gamma(B_{c}^*\to B_{c}\gamma)= (50\pm 30)$  eV, in our numerical evaluation, which leads to a large theoretical uncertainty.

 \begin{figure}[t]
\caption{The dependence of  form factors in the BSW model for $B_{c}\to{J/\psi}$ and $B_{c}^{*}\to{\eta_c}$ transitions  at $q^2=0$ on the parameter $\omega$. }
\begin{center}
\subfigure[]{\includegraphics[scale=0.8]{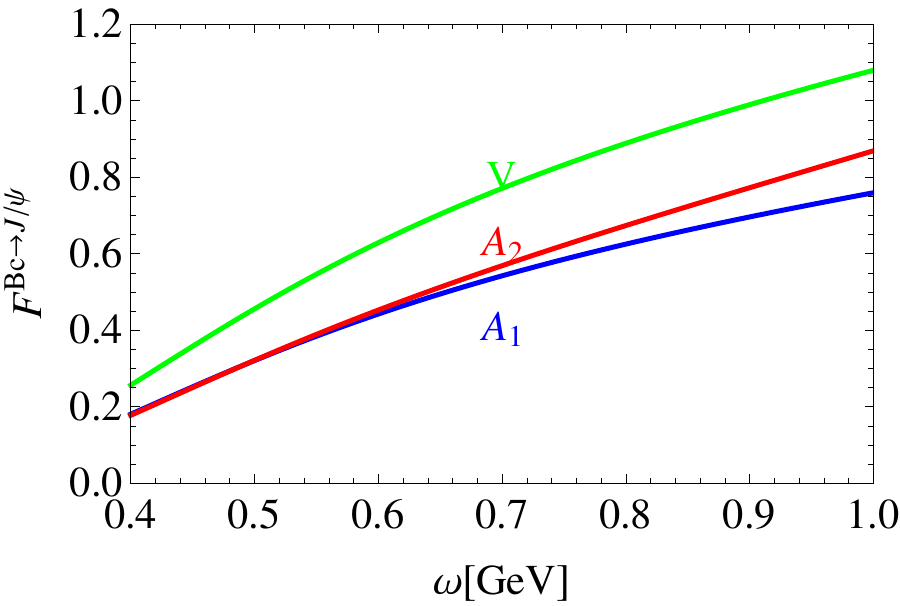}}\qquad\quad
\subfigure[]{\includegraphics[scale=0.8]{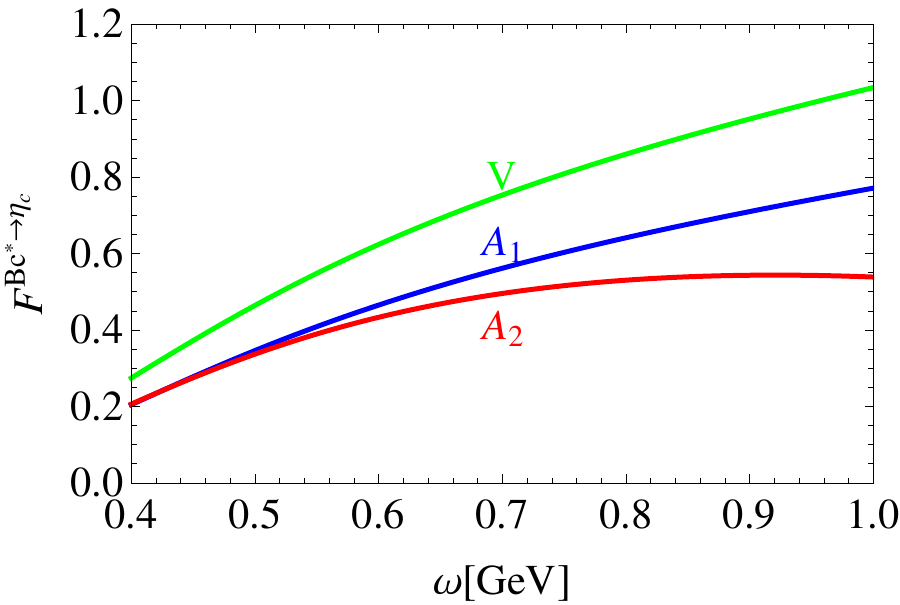}}
\end{center}
\label{fig:bswff}
\end{figure}
  \begin{table}[t]
   \caption{ The values of form factors at $q^2=0$ for $B_{c}\to{J/\psi}$ and $B_{c}^{*}\to{\eta_c}$ transitions in the BSW model with different values of $\omega$~(in units of GeV).}
   \label{tab:ff1}
   \begin{center}\setlength{\tabcolsep}{5pt}
   \begin{tabular}{ccccc}\hline\hline
    {Transition} & {Form factors} & $\omega=0.4$  & $\omega=0.6$
                                                 & $\omega=0.8$ \\ \hline
                                              &$A_{1}$  & 0.180  &0.442    & 0.624  \\
      {$B_{c}\to{J/\psi}$}       &$A_{2}$  & 0.177   &0.452   & 0.673\\
                                              &$V$         & 0.256   & 0.629   & 0.888 \\ \hline
                                              &$A_{1}$ &0.204   &0.465   &0.641  \\
    {$B_{c}^{*}\to{\eta_c}$}   &$A_{2}$   &0.206  &0.433   &0.530  \\
                                              &$V$         &0.274  &0.624    &0.860  \\ \hline\hline
    \end{tabular}
  \end{center}
  \end{table}
The form factors are the important inputs for evaluating the amplitude of non-leptonic $B^{(*)}_{c}$ decay. In the BSW model, the form factors are very sensitive to the parameter $\omega$, which can be easily seen from the Fig.~\ref{fig:bswff}. Their values change quite dramatically  when $\omega < 0.7$ GeV, but the changes become relatively slow when $\omega \geq 0.8\, {\rm  GeV }$. For the $D\to M$ and $B_{d,u}\to M $~($M$ denotes a light meson) transitions, the value $\omega\sim[0.4,0.5]\,{\rm GeV}$ is suggested in Refs.~\cite{Wirbel:1985ji,Bauer:1986bm,Bauer:1988fx} to fit data.
However, for the nonleptonic $B_c$ decays,  the authors of Refs.~\cite{Du:1988ws,Du:1991np} find that a relatively large $\omega \sim 0.8\,{\rm GeV}$ may be a reasonable choice.   Taking $\omega=0.4\,,0.6\,,0.8\,{\rm GeV}$, our results of form factors at $q^2=0$ for $B_{c}\to{J/\psi}$ and $B_{c}^{*}\to{\eta_c}$ transitions  are listed in  Table~\ref{tab:ff1}.

In the LFQM, the parameter $\beta$ in Eq.~(\ref{eq:wavelf1}) plays a similar role as $\omega$ in the BSW model.
Its value has been well determined. In this paper, we take  $\beta_{\eta_c}=0.814^{+0.092}_{-0.086}$GeV, $\beta_{J/\psi}=0.632^{+0.005}_{-0.005}$GeV, $\beta_{B_c}=0.890^{+0.075}_{-0.074}$GeV and $\beta_{B_c^*}=0.75^{+0.080}_{-0.080}$GeV suggested by Refs.~\cite{Cheng:2003sm,Wang:2008xt,Scora:1995ty}. Using these inputs, we obtain the numerical results of  form factors,
 \begin{eqnarray}
&&A_{1}^{B_{c}\to J/\psi}(0)=0.493^{+0.009}_{-0.012}\,,\quad A_{2}^{B_{c}\to J/\psi}(0)=0.458^{+0.014}_{-0.020}\,,\quad V^{B_{c}\to J/\psi}(0)=0.742^{+0.009}_{-0.009}\,;\\
&&A_{1}^{B_{c}^{*}\to \eta_{c}}(0)=0.513^{+0.032}_{-0.041}\,,\quad A_{2}^{B_{c}^{*}\to \eta_{c}}(0)=0.410^{+0.028}_{-0.043}\,,\quad V^{B_{c}^{*}\to \eta_{c}}(0)=0.823^{+0.053}_{-0.051}\,.
 \end{eqnarray}

Comparing these results with the ones listed in Table~\ref{tab:ff1}, it can be easily found that the predictions in the BSW model  with $\omega \sim [0.7,0.8]\,{\rm GeV}$  are  consistent with the ones in the LFQM.  As is mentioned above, such value is also suggested   for $B_c$ decays by Refs.~\cite{Du:1988ws,Du:1991np}. Therefore,
in this paper, we take $\omega=0.7\pm 0.1\,{\rm GeV}$.  Then, we can obtain the numerical results in the BSW model,
 \begin{eqnarray}
&&A_{1}^{B_{c}\to J/\psi}(0)=0.542^{+0.083}_{-0.100}\,,\quad A_{2}^{B_{c}\to J/\psi}(0)=0.568^{+0.105}_{-0.116}\,,\quad V^{B_{c}\to J/\psi}(0)=0.771^{+0.118}_{-0.142}\,;\\
&&A_{1}^{B_{c}^{*}\to \eta_{c}}(0)=0.561^{+0.080}_{-0.096}\,,\quad A_{2}^{B_{c}^{*}\to \eta_{c}}(0)=0.495^{+0.035}_{-0.062}\,,\quad V^{B_{c}^{*}\to \eta_{c}}(0)=0.753^{+0.107}_{-0.129}\,.
 \end{eqnarray}

 \begin{table}[t]
\caption{Theoretical prediction for the observables of ${B_{c}}\to J/\psi V$ decays
   in comparison with  the results obtained in the other model.
  }
\begin{center}\footnotesize  \setlength{\tabcolsep}{1pt}
\begin{tabular}{clccccc}
\hline\hline
 \multirow{2}{*}{ Obs.} & \multirow{2}{*}{ Decay mode}  &\multicolumn{2}{c}{this work}&\multicolumn{3}{c}{Refs.} \\
&&BSW model&LFQM& \cite{Ebert:2003cn,Ebert:2003wc} & \cite{Kar:2013fna} &\cite{prd9605451}\\\hline
  ${\cal B}\,[\times10^{-4}]$
 &${B_{c}}\to J/\psi K^{*-}$       &$1.7^{+0.1+0.9}_{-0.1-0.8} $
                                                    &$1.6^{+0.1+0.1}_{-0.1-0.1}$
                                                   &$1.0$&$1.0$&$2.2$ \\
&${B_{c}}\to J/\psi\rho^{-}$     &$30.0^{+1.0+15.6}_{-1.3-14.4}$
                                                  &$27.5^{+1.0+1.7}_{-1.2-1.9}$
                                                     &$16$&$18.7$~&$40$ \\  \hline
$f_L\,[\%]$
 &${B_{c}}\to J/\psi K^{*-}$      &$83.6^{+3.4}_{-5.7}$
                                                   &$84.8^{+0.5}_{-0.5}$
                                                   &$-$&$91.1$&$-$\\
 &${B_{c}}\to J/\psi\rho^-$      &$87.0^{+2.9}_{-4.9}$
                                                  &$88.0^{+0.5}_{-0.4} $
                                                   &$-$&$93.2$&$-$\\
\hline
$f_{\|}\,[\%]$
&${B_{c}}\to J/\psi K^{*-}$       &$13.6^{+3.5}_{-2.5}$
                                                   &$12.3^{+0.3}_{-0.4} $
                                                   &$-$&$6.5$&$-$\\
 &${B_{c}}\to J/\psi \rho^-$     &$ 10.7^{+3.1}_{-2.1}$
                                                  & $9.7^{+0.3}_{-0.3} $
                                                   &$-$&$4.9$&$-$\\
 \hline\hline
\end{tabular}
\end{center}
\label{tab:Br1}
\end{table}
 \begin{table}[t]
\caption{Theoretical prediction for the observables of $B_{c}^{*}\to \eta_{c}V$ decays.
 }
\begin{center}\footnotesize  \setlength{\tabcolsep}{1pt}
\begin{tabular}{llcc}
\hline\hline
 Obs. & Decay mode&BSW model   &LFQM \\\hline
{${\cal B}\,[\times10^{-9}]$}  &${B}_c^{*} \to {\eta_c} K^{*-}$    &$1.7^{+0.1+0.4+2.6}_{-0.1-0.4-0.6}$          &$1.4^{+0.1+0.1+2.1}_{-0.1-0.2-0.5}$\\
                                                &${B}_c^{*} \to {\eta_c}\rho^-$    &$30.2^{+0.1+7.1+45.2}_{-1.3-7.6-11.3}$     &$24.5^{+0.8+2.5+36.8}_{-1.1-3.1-9.2}$\\
\hline
$f_L\,[\%]$            &${B}_c^{*} \to {\eta_c} K^{*-}$      &$84.6^{+0.8}_{-1.0}$    &$83.1^{+0.6}_{-0.5}$ \\
                             &${B}_c^{*} \to {\eta_c}\rho^-$       &$88.0^{+0.7}_{-0.9}$    &$86.6^{+0.5}_{-0.6}$ \\\hline
$f_{\|}\,[\%]$        &${B}_c^{*} \to {\eta_c} K^{*-}$       &$12.7^{+1.0}_{-1.1}$    &$13.0^{+0.6}_{-0.6}$\\
                            &${B}_c^{*} \to {\eta_c}\rho^-$       &$10.0^{+0.8}_{-0.9}$    &$10.2^{+0.5}_{-0.5}$\\
\hline\hline
\end{tabular}
\end{center}
\label{tab:Br2}
\end{table}

  Using the values of inputs given above and the theoretical formula given in the last section, we then present our theoretical prediction and discussion for the ${B_{c}}\to J/\psi V$ and $B_{c}^{*}\to \eta_{c}V$ decays. Our numerical results for the observables at the scale of $\mu=m_b$ are summarized in Tables \ref{tab:Br1} and \ref{tab:Br2}.  For the branching fractions, the two errors in Table \ref{tab:Br1} and the first two errors in  Table \ref{tab:Br2} are caused by the inputs listed in Table \ref{tab:input} and the form factors, respectively; and the third error in  Table \ref{tab:Br2} is caused by the decay width of $B_c^*$. For the polarization fractions in Tables \ref{tab:Br1} and \ref{tab:Br2}, only the error induced by the   form factors are given because the errors induced by the other inputs are numerically negligible. The followings are some analyses and discussions:
\begin{itemize}
\item
Due to the hierarchy of the CKM matrix elements $|V_{ud}|>|V_{us}|$,
    there obviously exist the relations,
  \begin{eqnarray}
 {\cal B}(B_{c}\to J/\psi \rho^-) &> &{\cal B}(B_{c}\to J/\psi K^{*-})\,,\\
  {\cal B}(B_{c}^{*}\to \eta_{c}\rho^-) &>& {\cal B}(B_{c}^{*}\to \eta_{c}K^{*-})\,.
  \label{br1}
 \end{eqnarray}
Moreover, according to the magnitude of $|V_{ud}|\sim 1$ and $|V_{us}|\sim \lambda$,
it is expected that  $ {\cal B}(B_{c}\to J/\psi \rho^-)/{\cal B}(B_{c}\to J/\psi K^{*-})\approx {\cal B}(B_{c}^{*}\to \eta_{c}\rho^-)/{\cal B}(B_{c}^{*}\to \eta_{c}K^{*-})\approx 1/\lambda^2 \sim 20$, which is  required by the $SU(3)$ flavor symmetry. These relations can be easily tested from the results listed in Tables \ref{tab:Br1} and  \ref{tab:Br2}.

\item
The expected $B_c$ cross section at the LHC is at the level of $1\mu {\rm b}$~\cite{Brambilla:2010cs} and the integrated luminosity will reach up to $50\,{\rm fb^{-1}}$~(10 years) at LHCb~\cite{Bediaga:2012py}, so a very large number of $B_c$ will be produced and recorded. The estimation in  Ref.~\cite{Gouz:2002kk} shows that about $4.5\times10^{10}$  $B_{c}$ samples can be produced  per year at LHC. As it is clear from Table \ref{tab:Br1}, branching fractions of $B_{c}\to J/\psi \rho^-$ and $J/\psi K^{*-}$ decay modes are large enough for reliable measurements at LHCb in the future.

Assuming $\sigma(B_{c}^{*}) \sim 0.5 \mu{\rm b}$ and using the results in Table \ref{tab:Br2}, about 755 and 43
signal events are expected for $B_{c}^{*}\to \eta_{c}\rho^-$ and $\eta_{c}K^{*-}$ decays, respectively,  with total $50\,{\rm fb^{-1}}$ of data. However, after considering the reconstruction efficiency, these decay modes are not very easy to be observed by LHCb.



\item

In Table \ref{tab:Br1}, the predictions given  in the previous works~ \cite{Kar:2013fna,Ebert:2003cn,Ebert:2003wc,prd9605451} are also listed for comparison. There results are obtained with the form factors estimated in the relativistic quark model~\cite{Ebert:2003cn,Ebert:2003wc}, relativistic independent quark model~\cite{Kar:2013fna} and QCD sum rules~\cite{prd9605451}, respectively.
It can be found that our results~(central values)  for the branching fractions are a little larger than the ones in Refs.~\cite{Kar:2013fna,Ebert:2003cn,Ebert:2003wc}, but smaller than the ones in Ref.~\cite{prd9605451}.  These theoretical results are generally coincide with each other within errors.

The previous works aforementioned are  based on the naive factorization~(NF) approach, which results in $a_{1}^{0,\mp}=C_1+C_2/N_c\simeq 1.018$~(NF) at $\mu=m_b$. In this work, we find that $a_{1}^{0,-,+}$ are enhanced by a factor of about $(3.4,2.8,5.0)\%$ by vertex QCD corrections, and the total correction to the branching fraction is about $8\%$.

\item
In the previous works for the $b\to c$ induced two-body non-leptonic $B$ decays, the transverse contributions are usually neglected because they are power-suppressed relative to the longitudinal amplitude. For the decay modes considered in this paper,  it is expected that the polarization amplitudes satisfy relation  ${\cal A}_0: {\cal A}_-:{\cal A}_+\sim 1:\frac{2m_{V}m_{B_c}}{m_{B_c}^2-m_{J/\Psi}^2}:\frac{2m_{V}m_{J/\Psi}}{m_{B_c}^2-m_{J/\Psi}^2}$ for $B_c\to J/\Psi V$ decays and $1:\frac{2m_{V}m_{B_c^*}}{m_{B_c^*}^2-m_{\eta_c}^2}:\frac{2m_{V}m_{\eta_c}}{m_{B_c^*}^2-m_{\eta_c}^2}$ for ${B}_c^{*} \to {\eta_c} V$ decays, which can be easily obtained from Eqs.~(\ref{eq:H0B}), (\ref{eq:HmpB}) and Eqs.~(\ref{eq:H0Bstar}), (\ref{eq:HmpBstar}), respectively. Our results listed in Tables \ref{tab:Br1} and  \ref{tab:Br2} generally follow these expectations.  Despite of that, the transverse amplitudes present about $(10\sim20)\%$ contribution to the branching fraction, and therefore are numerically un-negligible.

 \item
From Tables \ref{tab:Br1} and  \ref{tab:Br2}, it can be found that the large theoretical errors for the branching fractions are mainly induced by the form factors. Beside, the underdetermined lifetime of $B^*_c$ meson also leads to large  uncertainties for $B^*_c$ decays. Fortunately, instead of evaluating the branching fractions directly, these theoretical uncertainties can be well controlled by evaluating their ratios. Numerically, using the form factors in the BSW model, we obtain
  \begin{eqnarray}
  R_{K^{*}/\rho}&\equiv&{\cal B}(B_{c}\to J/\psi  K^{*-})/{\cal B}(B_{c}\to J/\psi \rho^-) = 0.057\pm 0.004\,,\\
 R_{K^{*}/\rho}^{\prime}&\equiv&{\cal B}(B_{c}^{*}\to \eta_{c}  K^{*-})/{\cal B}(B_{c}^{*}\to \eta_{c} \rho^-) = 0.057\pm 0.004\,.
 \end{eqnarray}
 Moreover, the uncertainties can be further reduced for a given polarization component, for instance,  $R_{K^{*}/\rho}^{\lambda=0}=0.055\pm0.003$.

 Recently, the ratio  $R_{K/\pi} \equiv{\cal B}(B_{c}\to J/\psi  K^{-})/{\cal B}(B_{c}\to J/\psi \pi^{-}) =0.079\pm 0.007{\rm (stat.)}\pm 0.003{\rm (syst.)}$ has been measured by the LHCb collaboration~\cite{Aaij:2016tcz}. Using the data $f_{\pi}=130.3\,{\rm MeV}$ and $f_{K}=156.1\,{\rm MeV}$~\cite{Patrignani:2016xqp}, we can further obtain
   \begin{eqnarray}\label{data:piK}
  \widetilde{R}_{K/\pi}\equiv R_{K/\pi}\left(\frac{f_{\pi}}{f_{K}}\right)^2 = 0.055\pm 0.005\,,
 \end{eqnarray}
  which is expected to be equal to $\widetilde{R}_{K^{*}/\rho}^{\lambda=0}\equiv R_{K^{*}/\rho}^{\lambda=0}\left(\frac{f_{\rho}}{f_{K^{*}}}\right)^2$ approximately. It should be noted that the theoretical prediction for $\widetilde{R}_{K^{*}/\rho}^{\lambda=0}$ is independent of the decay constant, and therefore, the theoretical uncertainties can be further reduced.  Numerically, we find that our prediction $\widetilde{R}_{K^{*}/\rho}^{\lambda=0}=0.053\pm0.001$ are in good consistence with the experiment result, Eq.~(\ref{data:piK}), within $1\sigma$ error. The direct measurements on $R_{K^{*}/\rho}^{(\lambda=0)}$ and $\widetilde{R}_{K^{*}/\rho}^{(\lambda=0)}$ are required to confirm such consistence.

\end{itemize}

   \section{Summary}
  \label{sec04}

With the running and upgrading of the LHCb experiment, huge amounts of $B_c$ and $B_c^*$ mesons will be
produced, which provides us with a possibility of searching for their weak decays. In this paper, the nonleptonic two-body $B_{c}\to J/\psi V$ and $B_{c}^{*}\to\eta_{c}V$~($V=K^*\,,\rho$) decays are studied.
The NLO QCD corrections to the longitudinal and transverse amplitudes are evaluated within the framework of QCD factorization, and the transition form factors are calculated by using the Bauer-Stech-Wirbel model and light-front quark model. It is found that (i) the  NLO vertex contribution presents $\sim 8\%$ correction to the branching fraction; (ii) these decays are dominated by the longitudinal polarization, but the power-suppressed transverse corrections account for over $10\%$  of the whole contribution and therefore are un-negligible; (iii) the large theoretical uncertainties can be effectively controlled for some useful ratios, for instance, $R_{K^{*}/\rho}^{(\lambda=0)}$ and  $\widetilde{R}^{(\lambda=0)}_{K^{*}/\rho}$; and our prediction $\widetilde{R}_{K^{*}/\rho}^{\,\lambda=0}=0.053\pm0.001$ are in good consistence with the data $\widetilde{R}_{K/\pi}= 0.055\pm 0.005$ within $1\sigma$ error.
 Some of the results and findings will be tested by the LHCb experiment in the near future.

  \section*{Acknowledgments}
 This work is supported by the National Natural Science Foundation of China (Grant No. 11475055), Foundation for the Author of National Excellent Doctoral Dissertation of China (Grant No. 201317), Program for Innovative Research Team in University of Henan Province and the Excellent Youth Foundation of HNNU.
 
  \end{document}